\documentclass[conference]{IEEEtran}
\usepackage{cite}

\usepackage[usenames,dvipsnames,svgnames,table]{xcolor}

\ifCLASSINFOpdf
   \usepackage[pdftex]{graphicx}
\else
\fi
\usepackage{amsmath}
\DeclareMathOperator*{\POPCOUNT}{POPCOUNT}

\usepackage{array}

\begin{document}
\title{Outstanding Bit Error Tolerance  of  \\ Resistive RAM-Based Binarized Neural Networks}
\author{\IEEEauthorblockN{
T. Hirtzlin\IEEEauthorrefmark{1}, 
M. Bocquet\IEEEauthorrefmark{2}, 
J.-O. Klein\IEEEauthorrefmark{1}, 
E. Nowak\IEEEauthorrefmark{3}, 
E. Vianello\IEEEauthorrefmark{3},
J.-M. Portal\IEEEauthorrefmark{2} 
and 
D. Querlioz\IEEEauthorrefmark{1}}
\IEEEauthorblockA{\IEEEauthorrefmark{1}C2N, CNRS, Univ Paris-Sud, Universit\'e Paris-Saclay, 91405 Orsay cedex, France\\ 
Email: damien.querlioz@u-psud.fr}
\IEEEauthorblockA{\IEEEauthorrefmark{2}Institut Mat\'eriaux Micro\'electronique Nanosciences de Provence, Univ. Aix-Marseille et Toulon, CNRS, France}
\IEEEauthorblockA{\IEEEauthorrefmark{3}CEA, LETI, Grenoble, France.}
}
\IEEEoverridecommandlockouts
\maketitle
\begin{abstract}
Resistive random access memories (RRAM) are novel nonvolatile memory technologies, 
which can be embedded at the core of CMOS,
and which could be ideal for the in-memory implementation of deep neural networks.
A particularly exciting vision is using them for implementing Binarized Neural Networks (BNNs), a class of deep neural networks with a highly reduced memory footprint.
The challenge of resistive memory, however, is that they are prone to device variation, which can lead to bit errors.
In this work we show that BNNs can tolerate these bit errors to an outstanding level, through simulations of networks on the MNIST and CIFAR10 tasks.
If a standard BNN is used, up to $10^{-4}$ bit error rate can be tolerated with little impact on recognition performance on both MNIST and CIFAR10.
We then show that by adapting the training procedure to the fact that the BNN will be operated on error-prone hardware, this tolerance can be extended to a bit error rate of $4 \times 10^{-2}$.
The requirements for RRAM are therefore a lot less stringent for BNNs than more traditional applications. We show, based on experimental measurements on a RRAM $HfO_2$ technology, that this result can allow reduce  RRAM programming energy by a factor $30$.
%\textcolor{red}{Replace by quantitative statement.}
\end{abstract}


% \fi
\IEEEpeerreviewmaketitle

\section{Introduction}

Deep neural networks have made fantastic achievements in recent years \cite{lecun2015deep}.
Unfortunately, their high energy consumption 
limits their use in embedded applications \cite{editorial_big_2018,suleiman2017towards}. 
The in- or near-memory hardware implementation of deep neural networks is widely seen as a solution \cite{editorial_big_2018,ielmini2018memory,yu2018neuro,querlioz2015bioinspired}, as such implementations could avoid  entirely the energy cost of the von Neumann bottleneck. 
This is especially true with the emergence of novel memory technologies such as resistive random access memory (RRAM), phase change memory or spin torque magnetoresistive memory, which have made tremendous progress in recent years \cite{yu2018neuro}. These technologies provide low-area, fast and non-volatile memory cells, which can be embedded at the core of CMOS.

A challenge of the in-memory implementation of neural networks is the high amount of required memory. 
Binarized Neural Networks (BNNs)  have recently emerged as a possibly ideal solution \cite{courbariaux2016binarized,rastegari2016xnor}. 
These neural networks, where the synaptic weights as well as neuron activation are binary values,
can achieve near state-of-the-art performance on image recognition tasks, while using only a fraction of the memory used by conventional neural networks \cite{courbariaux2016binarized,rastegari2016xnor}.
BNNs are therefore an excellent candidate for hardware implementation where the synaptic weights are stored in RRAMs \cite{yu2018neuro,bocquet2018}.

Nevertheless, despite their outstanding qualities, emerging memories are  prone to device variation \cite{ly2018role,ielmini2018memory}, which can cause bit errors. 
In conventional applications, this is solved either by relying on error correcting codes \cite{zhang2017130nm},
or by programming memory cells with high energy pulses that lead to more reliable programming \cite{ly2018role}.
In this work, based on the experimental measurements of RRAM cells and system level simulations, we investigate the impact of bit errors on in-memory BNNs.
We find that BNNs can exhibit outstanding error tolerance, allowing us to avoid these traditional techniques for dealing with RRAM variability.

After presenting the background of the work (section~\ref{sec:background}):

\begin{itemize}
\item We show on several tasks that BNNs have  a high tolerance to RRAM bit error rate (section~\ref{sec:standard_training}).
\item We show that this tolerance can be extended to outstanding levels, if the training process of the BNNs takes into account the fact the neural networks is going to be operated on error-prone hardware (section~\ref{sec:adapted_training}).
\end{itemize}

%%%%%%%%%%%%%%%%%%%%%%%%%%%%%%%%%%%%%%%%%%%%%

\section{Background}
\label{sec:background}

Binarized Neural Networks are simplifications of conventional neural networks, where synaptic weights, as well as neuron activation values, assume binary values ($+1$ or $-1$) instead of real values.
In these conditions, the equation for neuron activation $A$:
\begin{equation}
    A = f(\sum_i W_iX_i),
    \label{eq:activ_real}
\end{equation}
where $X_i$ are the neuron inputs,  $W_i$ the corresponding synaptic weights and $f$ the non-linear activation function, is considerably simplified
\begin{equation}
    A = \POPCOUNT_i(XNOR( W_i,X_i))>T,
        \label{eq:activ_BNN}
\end{equation}
where $\POPCOUNT$ is the  function that counts the number of $1s$, and $T$ is a learned threshold. 

During training, synaptic weights  also assume real weights. The binary weights, equal to  the sign of the real weight, are used in both the forward and backward passes, while the real weights are updated by the learning rule \cite{courbariaux2016binarized}. Once training is done, the real weights are no longer needed. 
BNNs are therefore extremely attractive for hardware implementation 
of inference operation  \cite{ando2017brein}
due to their particularly low memory requirement (one bit per synapse),
and due to to the fact that resource-hungry real multiplications in eq.~(\ref{eq:activ_real}) are replaced by simple binary XNOR gates in eq.~(\ref{eq:activ_BNN}).
%Due to their low memory requirements, and reliance on simple operations, BNNs are outstanding candidates for in-memory computing. 
An optimal implementation would be with RRAM
\cite{sun2018fully,sun2018xnor,tang2017binary,yu2018neuro}, as RRAM cells are much more compact than SRAM, and yet non-volatile.

%It should be noted that in BNNs, only inference is binarized. 
%Training involves real valued weights and real multiplication \cite{courbariaux2016binarized,rastegari2016xnor}.
%BNNs are therefore ideal for inference-only hardware, 
%pretrained in a classical fashion on Graphical Processing Units (GPUs).

\begin{figure}[htbp]
	\centering
	\includegraphics[width=2in]{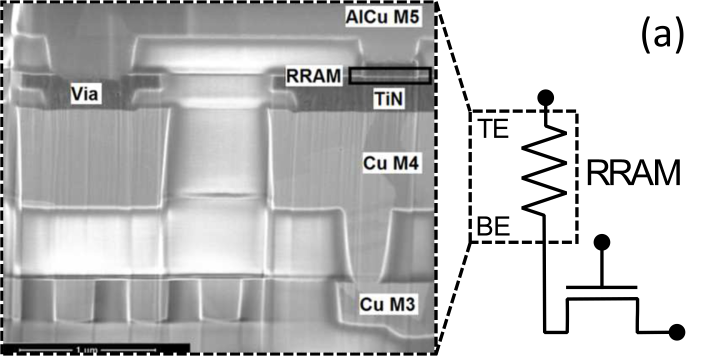}
	\vspace{0.5cm}
	\includegraphics[width=3.5in]{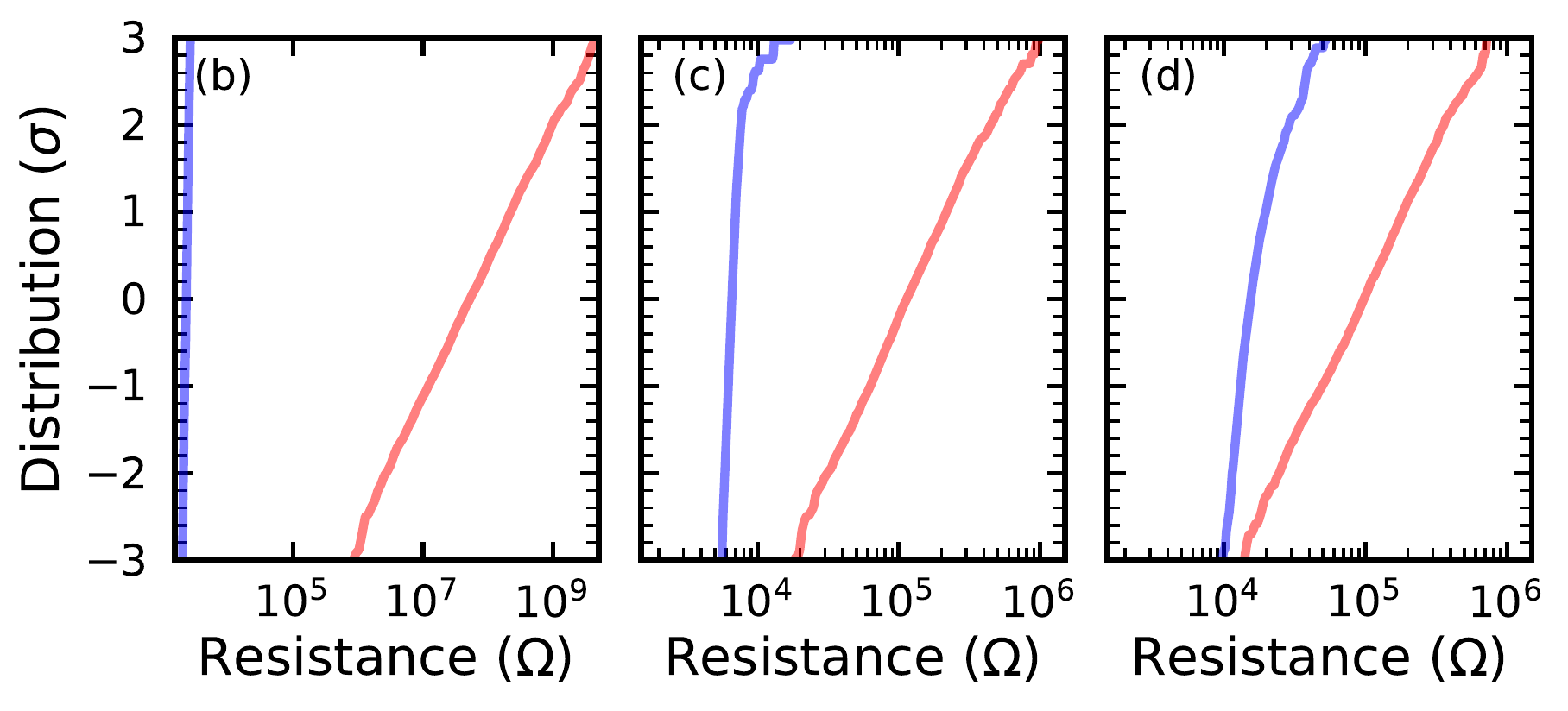}
	\caption{(a) Transmission electron microscopy image of a RRAM cell embedded in a CMOS process. (b-c) Distribution of the resistance states of RRAM cells in a kilobit array programmed in high resistance state (HRS, red line) and low resistance state (LRS, blue line). The RRAM array is programmed  with (b) very strong, (c) strong and (d) weak programming conditions. }
	\label{fig:RRAM}
\end{figure}

\begin{table}[htbp]
\caption{RRAM programming conditions of Fig.~\ref{fig:RRAM}(b-d). %\textcolor{red}{Rajouter Energie?}
}
\begin{center}
\begin{tabular}{|c|c|c|c|}
\hline
\textbf{Programming condition} &
\textbf{Very strong}
& \textbf{Strong} & \textbf{Weak} \\
\hline
SET compliance current & 
$600\mu A$ &
$55\mu A$ &
$20\mu A$ \\
\hline
RESET voltage & 
$2.5V$ &
$2.5V$ &
$1.5V$ \\
\hline
Programming time & $100 ns$& $100 ns$& $100 ns$ \\
\hline
Bit error rate & 
 $<10^{-6}$ &
$9.7 \times 10^{-5}$ &
 $3.3 \times 10^{-2}$ \\
\hline
Programming energy  & 
 $120/150pJ$ &
$11/14pJ$ &
 $4/5pJ$ \\
(SET/RESET) & & &\\
\hline
Cyclability & 
$100$ &
$>10,000$ &
$>10^6$ \\
\hline

\end{tabular}
\end{center}
\label{tab:programming_conditions}
\end{table}

Nevertheless, the challenge of using  RRAMs for in-memory computing is their device variation.
Fig.~\ref{fig:RRAM}(a) presents the transmission electron microscopy image of one of our  $HfO_2$-based RRAM cell, integrated
in the backend-of-line of a full CMOS process, on top of
the fourth layer of metal,
using the same process as
 \cite{grossi2016fundamental}. 
  These memory cells can be programmed either in Low Resistance State (LRS, meaning $0$) or High Resistance State (HRS, meaning $1$).
 Figs.~\ref{fig:RRAM}(b-d) show programming statistics of kilobit arrays of such memory cells.
Table~\ref{tab:programming_conditions} summarizes the programming conditions used in Figs.~\ref{fig:RRAM}(b-d), as well as the corresponding RRAM properties: bit error rate, and cyclability (number of times RRAM devices can be programmed before definitive failure). 
 
 %\textcolor{red}{Add programming conditions also in body text.}
 In Fig.~\ref{fig:RRAM}(b), the devices are programmed with ``very strong'' programming conditions (SET compliance current of $600 \mu A$, RESET voltage of $2.5V$, programming time $100ns$). These conditions  consume high programming energy, and also lead to device aging, causing low endurance: the RRAM devices cannot be programmed more than 100 times.
  It is seen in Fig.~\ref{fig:RRAM}(b) that, despite device variation, the distribution of resistance of the LRS and HRS do not overlap at $3\sigma$, leading to a bit error rate lower than $10^{-6}$.

  In Fig.~\ref{fig:RRAM}(c), the devices are programmed with ``strong'' programming conditions  (SET compliance current of $55 \mu A$, RESET voltage of $2.5V$, programming time $100ns$). These conditions  consume $11$ times less  programming energy than the previous ones, and also lead to less device aging. The devices can be programmed more than $10,000$ times.
 The distribution of resistance of the LRS and HRS do not overlap at $3\sigma$, leading to a bit error rate of $9.7 \times 10^{-5}$.

  In contrast, in Fig.~\ref{fig:RRAM}(d), the devices are programmed with ``weak'' programming conditions  (SET compliance current of $20 \mu A$, RESET voltage of $1.5V$, programming time $100ns$). These conditions  consume thirty times less programming energy than the very strong ones, and have very reduced device  aging: they can be programmed millions of times 
  The distribution of resistance of the LRS and HRS do  overlap significantly, leading to a high bit error rate of $3.3 \times 10^{-2}$.
  
  In applications, this issue of device variability can be dealt with several strategies. Either, we can use very strong programming conditions, causing high energy consumption, larger cell area (as the transistor associated with RRAM cells need to be able to drive high currents), and low endurance. Alternatively, we can rely on error correcting codes \cite{zhang2017130nm}, causing a very significant overhead to implement decoding circuits  \cite{gregori2003chip}.
  Other strategies such as write termination \cite{chang201419,su2017462gops} and adaptive programming \cite{sassine2018sub} can also reduce the impact of device variability, but with the cost of strong area overhead. 
  Now, we look how BNNs can deal with the issue of device variation in a simpler fashion.
This is especially important as some strategies proposed to enhance bit error tolerance in synaptic weights such as the reliance on sign-magnitude representation \cite{whatmough2018dnn} do not apply to BNNs.
%\textcolor{red}{Techniques for bit error toleranceof neural networks such as Whatmough and Reagen cannot be used with BNNs.}

%%%%%%%%%%%%%%%%%%%%%%%%%%%%%%%%%%%%%%%%%%%

\section{Bit Error Tolerance with Traditional Training Method}
\label{sec:standard_training}

\begin{figure}[htbp]
	\centering
	\includegraphics[width=3in]{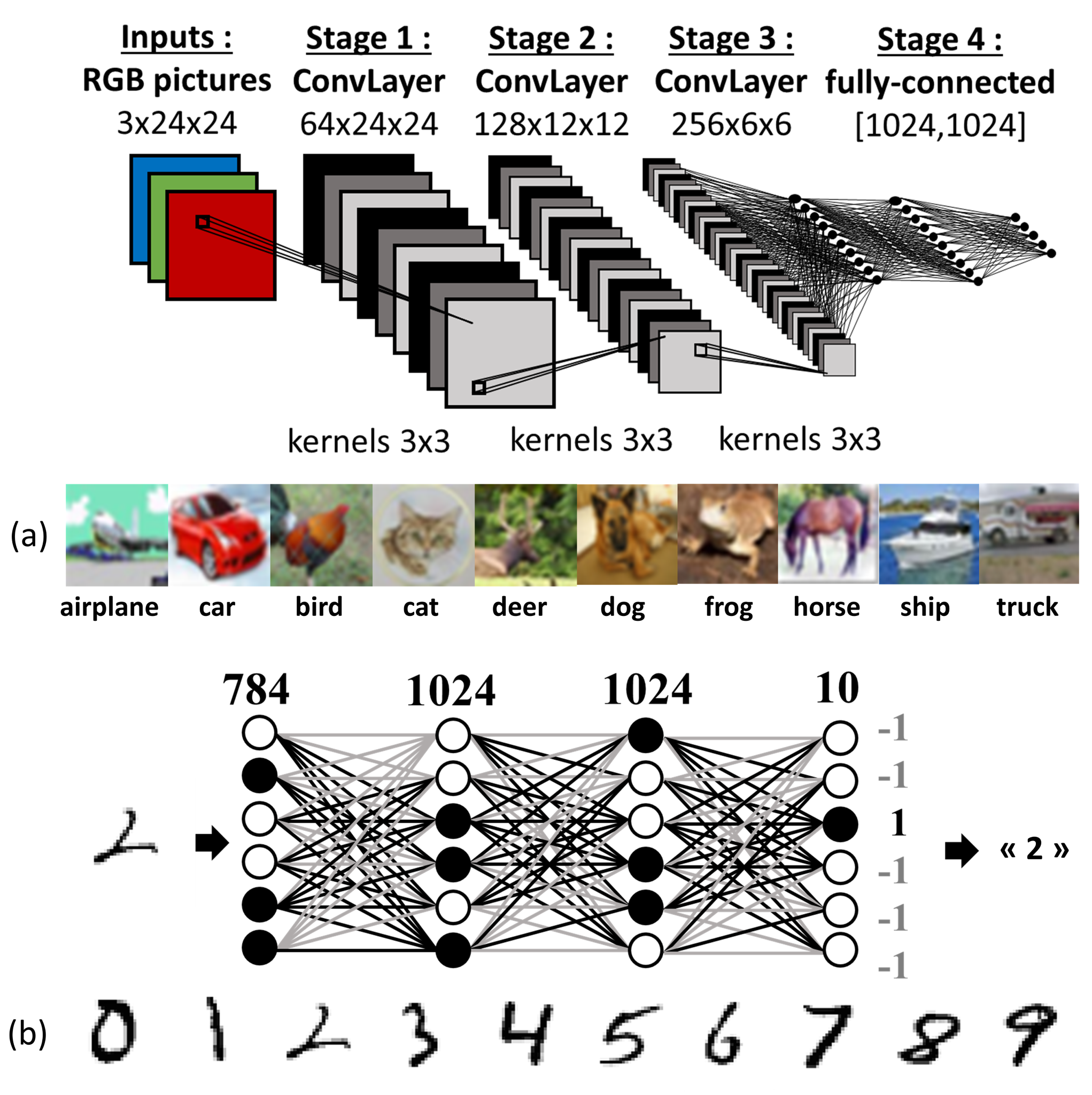}
	\caption{Binarized Neural Networks considered within this paper. 
	(a) Fully connected network for MNIST handwritten digit recognition.
	(b) Convolutional neural network for CIFAR10 image recognition.	}
	\label{fig:BNN}
\end{figure}

\begin{figure}[htbp]
	\centering
	\includegraphics[width=3in]{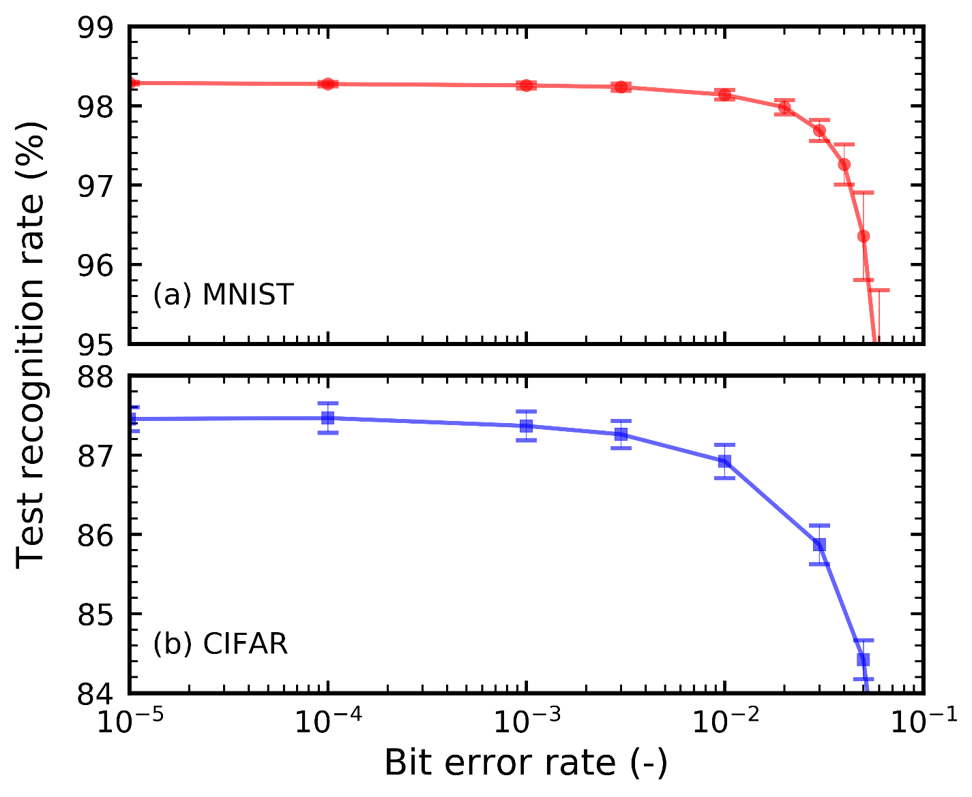}
	\caption{Recognition rate on the test dataset of (a) the fully connected neural network for MNIST and (b) the convolutional neural network for CIFAR10, as a function of the bit error rate over the weights during inference. 
	The neural networks have been trained in a standard fashion, without weight errors.}
	\label{fig:MNIST_CIFAR_wolearning}
\end{figure}

To investigate the impact of bit errors due to RRAM device variation in BNNs with synaptic weights stored in RRAMs, we performed multiple BNNs simulations.
We first consider a fully connected neural network
{with two hidden layers of 4096 neurons, a softmax cross entropy loss function and dropout}, illustrated in Fig.~\ref{fig:BNN}.
We trained this neural network on a Nvidia Tesla V100 GPU to perform the traditional MNIST handwritten character recognition task  \cite{lecun1998gradient}
{for 1000 epochs}.
%\textcolor{red}{More on techniques.}
The BNNs were simulated in the python language using the deep learning Tensorflow framework, and an identical training procedure than \cite{courbariaux2016binarized}.
%Bit errors on the weights can be added during the training and testing phase. If bit errors are introduced during training, different errors are introduced at each minibatch, to enhance the resilience of the network. If errors are introduced at test time, the errors are introduced at the beginning before testing, and kept the same during the whole test process, which mimics programming errors in hardware. The tests are repeated ten times with different bit errors. On the graph, the accuracy point is the mean of these ten runs, and the error bar is one standard deviation.
Without any bit error, this network achieves a {$1.6\%$} error rate on the test dataset.

We then test the neural network again, but introducing random bit errors on the synaptic weights, i.e randomly flipping a fraction $+1$ weights to $-1$, and of $-1$ weights to $+1$. 
Fig~\ref{fig:MNIST_CIFAR_wolearning}(a) shows the test recognition rate on MNIST as the function of the weight bit error rate.
We see that bit error rate as high as $10^{-3}$ does not affect the recognition rate.
With an error rate of $10^{-2}$ the recognition rate is only slightly reduced from {$1.6\%$} to {$1.8\%$}.

To check that this surprising result is not due to the simplicity of the considered neural network, we also studied a deep convolutional neural network, trained on the much more difficult image recognition CIFAR10 task \cite{krizhevsky2009learning}. 
The structure of the neural network is presented in Fig.~\ref{fig:BNN}.
Without bit errors, the test recognition rate is $87.6\%$.
All simulations were performed with the Tensorflow deep learning framework \cite{abadi2016tensorflow}.

We then perform the simulations including errors.
Fig~\ref{fig:MNIST_CIFAR_wolearning}(b) shows the test recognition rate on CIFAR10 as the function of the weight bit error rate.
Overall, the system is only slightly less robust than in the MNIST case.
With an error rate of $10^{-3}$ the recognition rate is reduced from {$87.6\%$} to {$87.4\%$}.
With an error rate of $10^{-2}$ it drops  to {$86.9\%$}.

We have seen in section~\ref{sec:background} that in RRAMs, the choice of programming conditions determines the bit error rate. 
As we only need ensure a bit error rate of $10^{-4}$ to avoid any recognition rate degradation in both MNIST and CIFAR10 tasks, we therefore do not need to use ``very strong'' programming conditions of Fig.~\ref{fig:RRAM} (b) and Table~\ref{tab:programming_conditions}, but can use the strong conditions. 
This saves programming energy, cell area and enhance the RRAM endurance, with regards to more conventional applications that would require the reliability of the very strong programming conditions.

%%%%%%%%%%%%%%%%%%%%%%%%%%%%%%%%%

\section{Adapting the Training Method can Extend the Bit Error Tolerance}
\label{sec:adapted_training}

We now show that the already high robustness seen in Fig.~\ref{fig:MNIST_CIFAR_wolearning} can be further enhanced if an appropriate training method is used.
For this purpose, we retrain the BNNs, but this time including bit errors \textit{during the training process}, and not only during the testing phase.
This way, the training process takes into account the fact that the BNN will be implemented on error-prone RRAM-based systems.
The devices subject to errors are chosen independently in training and testing phase:
the training phase assumes that the RRAM-based systems will have errors, but does not know which devices will be affected.

 More precisely, to train the neural network, we added errors at each iteration. The forward-pass is computed with errors on weights, e.g. {$10\%$} are changed from the original weights $W$ to $W_{error}$. During the backward-pass, we reuse the same value of weights $W_{error}$ instead of $W$ to backpropagate  through the whole depth of the neural network.

\begin{figure}[htbp]
	\centering
	\includegraphics[width=3.0in]{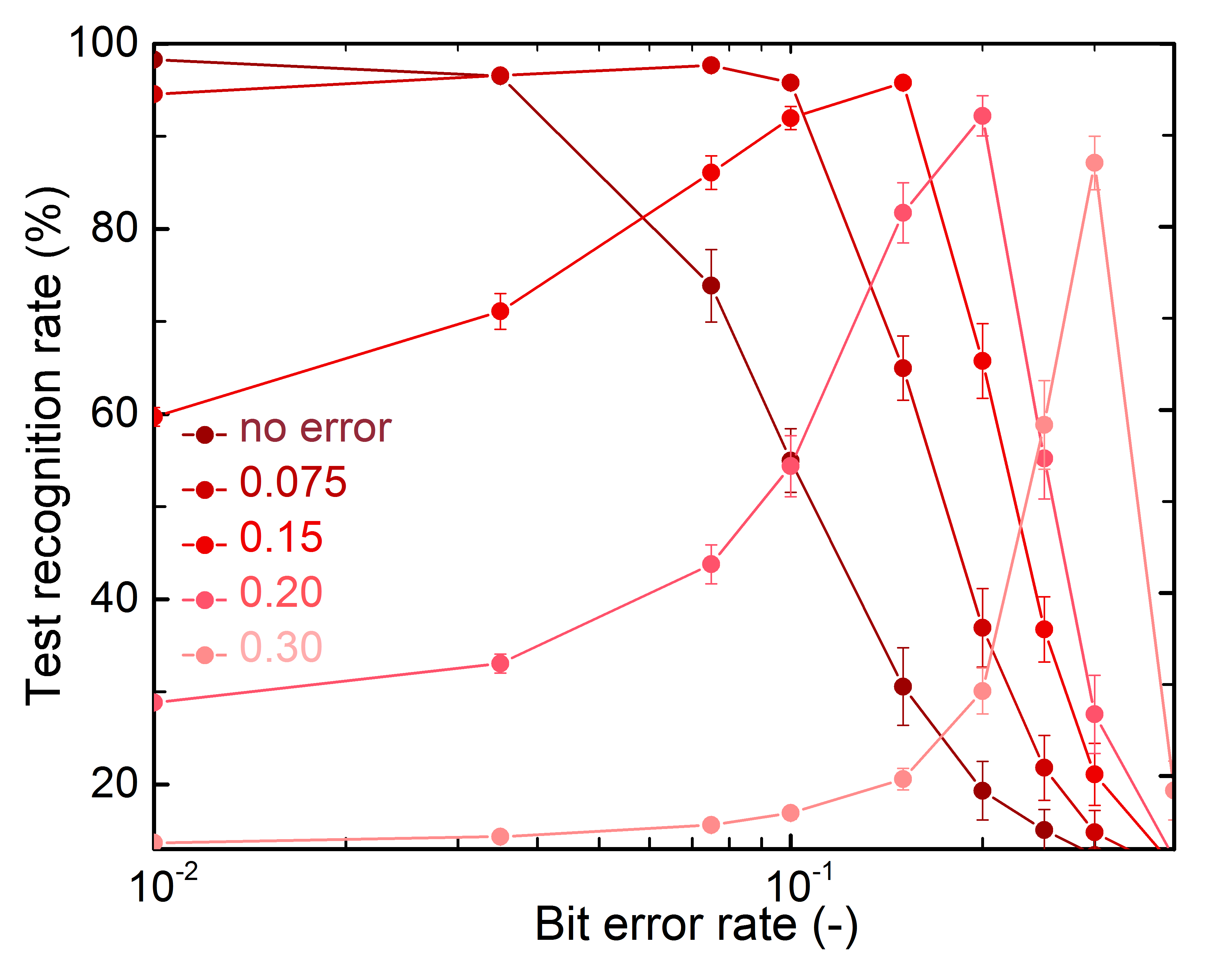}
	\caption{Recognition rate on the test dataset of the fully connected neural network for MNIST as a function of the bit error rate over the weights during inference. Dark red curve: no weight error considered during training. Other curves: the adapted training was used, each curve corresponding to a different bit error rate on the weights during training.	}
	\label{fig:MNISTwlearning}
\end{figure}

We now check if this training approach leads to more robust neural networks.
We performed the adapted training procedure with different error rates during training.
Fig.~\ref{fig:MNISTwlearning} shows the test error rate on the MNIST task, as a function of bit error rate during testing, for several error rates during training.
We see that these curves exhibit a maximum when error rate during testing matches the error rate during training: the network indeed finds a structure particularly adapted to the number of bit errors.
The recognition rate  at this maximum can be very high.
For a bit error rate of $0.15$, the recognition rate is {$95.9\%$}. If errors had not been taken into account during training (dark red line), recognition rate would only be {$30.4\%$}. 
This is quite astonishing that we can get the neural network to function so well, with $15\%$ of incorrect weights!

The fact that the recognition rate exhibits a maximum  as a function of bit error rate at test time seems counter intuitive. This can however be understood due to the particular properties of BNNs.
In these networks, the value of the neurons is binarized, following eq.~(\ref{eq:activ_BNN}). 
Having or not having bit errors in the weight $W_i$ can shift the mean value of $\POPCOUNT_i(XNOR( W_i,X_i))$.
This effect is very strong in the first layer of the neural network, because in MNIST, the input pixels are mostly binary (black or white).
Because of this, there can be a situation where when the bit error rate is shifted, the $\POPCOUNT$ value is always below threshold, or always above threshold $T$. Then, the neuron always has the same value and is useless for classification. For example, for a training bit error rate of $0.3$, and no bit error at test time, $48\%$ of the first layer neurons are in this situation. Whereas only $13\%$ of these neurons are in this situation if the $0.3$ bit error rate is used both during training and testing.

\begin{figure}[htbp]
	\centering
	\includegraphics[width=3.0in]{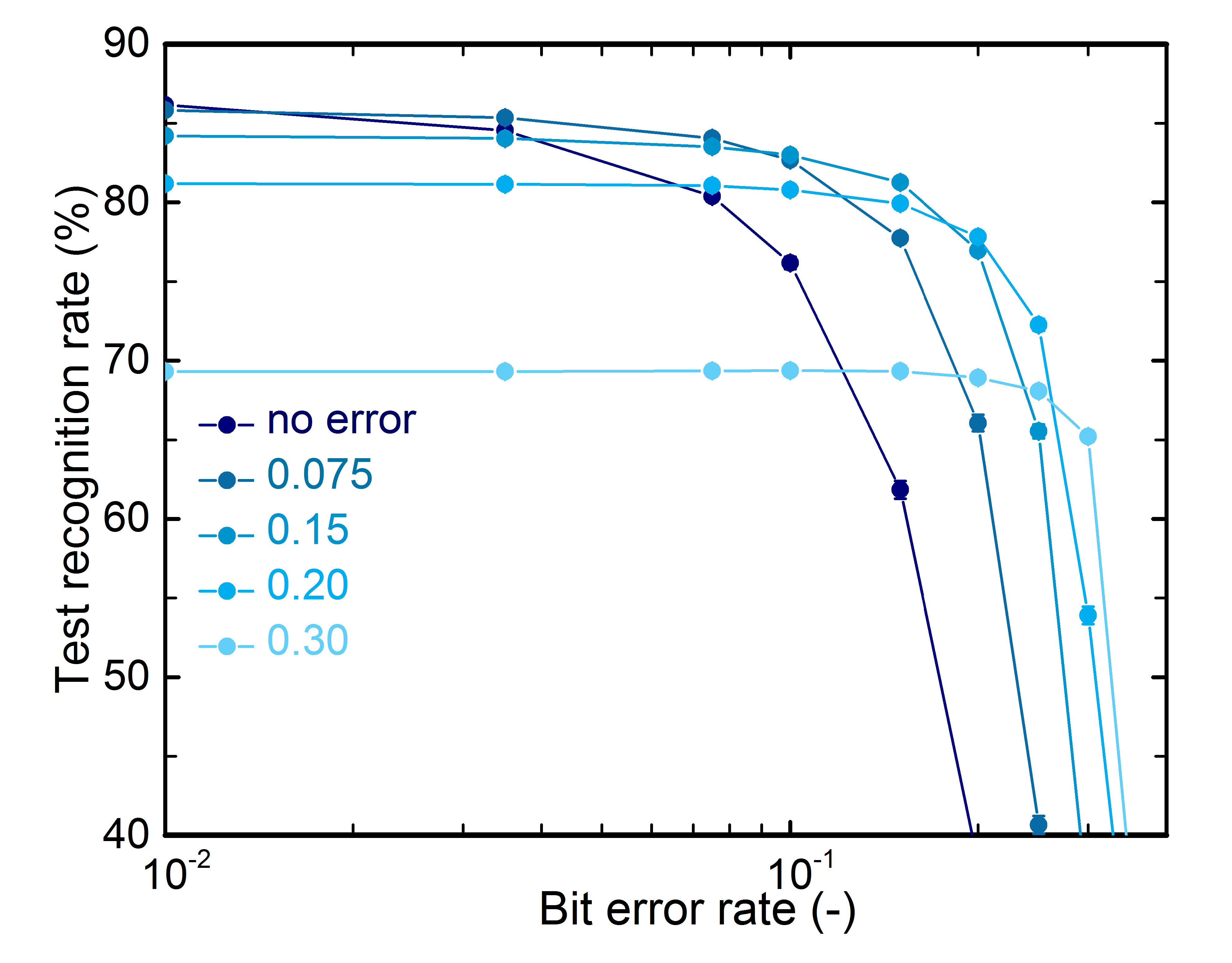}
	\caption{Recognition rate on the test dataset of the convolutional neural network for CIFAR10  as a function of the bit error rate over the weights during inference. Navy curve: no weight error considered during training. Other curves: the adapted training was used, each curve corresponding to a different bit error rate on the weights during training.}
	\label{fig:CIFARwlearning}
\end{figure}

Fig.~\ref{fig:CIFARwlearning} shows the results of the same study on the CIFAR10 dataset.
The results look slightly different than the MNIST case, 
in that the curves do not exhibit maxima: if the network has been trained with a given bit error rate, having less bit errors during testing is not detrimental.

This  can be explained by going back to the principle of BNNs. We observed that when presenting CIFAR10, adding bit errors does not shift significantly the mean values of  $\POPCOUNT_i(XNOR( W_i,X_i))$. The difference with MNIST comes from the fact that MNIST images are mostly black and white, whereas the CIFAR10 images feature rich colors.
Because of that, the error-induced shift of the mean value of $\POPCOUNT_i(XNOR( W_i,X_i))$ is significant in the case of the first layer of a neural network during MNIST, but much less significant on CIFAR10.

Once again, using this procedure, extremely high amount of bit errors can be tolerated. Bit error rates up to $4\times 10^{-2}$ do not affect the recognition rate.
For a bit error rate of $0.15$, the recognition rate is {$81.5\%$}, instead of  {$62.2\%$} if errors had not been taken into account during training (navy line). 

When using RRAM devices, using the adapted training procedure therefore allows us using directly the ``weak'' programming conditions of Fig.~\ref{fig:RRAM} and Table~\ref{tab:programming_conditions}, despite its high bit error rate.
This can allow us to benefit
from its low programming energy, cell area and high endurance of more than one million cycles.

%%%%%%%%%%%%%%%%%%%%%%%%%%%%%%%

%\begin{table}[htbp]
%\caption{RRAM programming conditions of Fig.~\ref{fig:RRAM}(b-d). 
%}
%\begin{center}
%\begin{tabular}{|c|c|c|c|} \hline
%\textbf{Programming condition} & \textbf{Very strong}& \textbf{Strong} & \textbf{Weak} \\
%\hline
%Standard Training & & & \\
%MNIST & $98.5\%$ & $98.4\%$  & \textcolor{red}{$96.5\%$}  \\
%CIFAR10 & $87.6\%$ &$87.5\%$  &\textcolor{red}{$85.6\%$}  \\
%\hline
%Adapted Training & & & \\ MNIST & $97.9\%$ &$97.9\%$  & \textcolor{blue}{$98.4\%$}  \\
%CIFAR10 & $87.5\%$ &$87.4\%$  &\textcolor{blue}{$87.3\%$}  \\
%\hline
%\end{tabular}
%\end{center}
%\label{tab:rec_rates}
%\end{table}

%%%%%%%%%%%%%%%%%%%%%%%%%

\section{Conclusion}

The in-memory implementation of Binarized Neural Networks (BNNs) with emerging memories such as RRAM is an exciting road for ultralow energy embedded artificial intelligence.
A typical challenge of emerging memories is their device variation, which can lead to bit errors.
In this work, we have seen that BNNs can tolerate a very high number of bit errors.
The BNN is trained in a conventional manner, up to $10^{-4}$ bit error rate on the synaptic weights can be tolerated on both MNIST and CIFAR task.
If the BNN, was trained with a specific procedure taking into account the fact that it will be operated on error-prone hardware, up $4 \times 10^{-2}$ bit error rate can be tolerated!

These results can have strong impact on emerging memory design and optimization toward neural network implementation.
If one accepts to increase device variability and bit error rate,
we saw that is is possible to increase the programming energy efficiency by a factor of thirty, decrease the area and increase the endurance of RRAM cells to more than one million cycles.
%\textcolor{red}{Replace by quantitative statement.}
This can significantly enhance their benefits.
The results of this paper also highlight the robustness of neural networks, even in the highly digital BNN form, and their adaptability to the constrains of hardware.

%rajouter apres acceptation
\section*{Acknowledgment}
This work was supported by the European Research Council Starting Grant NANOINFER (715872).

\bibliography{IEEEabrv,Mabibliotheque}

\begin{thebibliography}{10}
\providecommand{\url}[1]{#1}
\csname url@samestyle\endcsname
\providecommand{\newblock}{\relax}
\providecommand{\bibinfo}[2]{#2}
\providecommand{\BIBentrySTDinterwordspacing}{\spaceskip=0pt\relax}
\providecommand{\BIBentryALTinterwordstretchfactor}{4}
\providecommand{\BIBentryALTinterwordspacing}{\spaceskip=\fontdimen2\font plus
\BIBentryALTinterwordstretchfactor\fontdimen3\font minus
  \fontdimen4\font\relax}
\providecommand{\BIBforeignlanguage}[2]{{%
\expandafter\ifx\csname l@#1\endcsname\relax
\typeout{** WARNING: IEEEtran.bst: No hyphenation pattern has been}%
\typeout{** loaded for the language `#1'. Using the pattern for}%
\typeout{** the default language instead.}%
\else
\language=\csname l@#1\endcsname
\fi
#2}}
\providecommand{\BIBdecl}{\relax}
\BIBdecl

\bibitem{lecun2015deep}
Y.~LeCun, Y.~Bengio, and G.~Hinton, ``Deep learning,'' \emph{Nature}, vol. 521,
  no. 7553, p. 436, 2015.

\bibitem{editorial_big_2018}
Editorial, ``\BIBforeignlanguage{EN}{Big data needs a hardware revolution},''
  \emph{\BIBforeignlanguage{EN}{Nature}}, vol. 554, no. 7691, p. 145, Feb.
  2018.

\bibitem{suleiman2017towards}
A.~Suleiman, Y.-H. Chen, J.~Emer, and V.~Sze, ``Towards closing the energy gap
  between hog and cnn features for embedded vision,'' in \emph{Proc.
  ISCAS}.\hskip 1em plus 0.5em minus 0.4em\relax IEEE, 2017, pp. 1--4.

\bibitem{ielmini2018memory}
D.~Ielmini and H.-S.~P. Wong, ``In-memory computing with resistive switching
  devices,'' \emph{Nature Electronics}, vol.~1, no.~6, p. 333, 2018.

\bibitem{yu2018neuro}
S.~Yu, ``Neuro-inspired computing with emerging nonvolatile memorys,''
  \emph{Proc. IEEE}, vol. 106, no.~2, pp. 260--285, 2018.

\bibitem{querlioz2015bioinspired}
D.~Querlioz, O.~Bichler, A.~F. Vincent, and C.~Gamrat, ``Bioinspired
  programming of memory devices for implementing an inference engine,''
  \emph{Proc. IEEE}, vol. 103, no.~8, pp. 1398--1416, 2015.

\bibitem{courbariaux2016binarized}
M.~Courbariaux, I.~Hubara, D.~Soudry, R.~El-Yaniv, and Y.~Bengio, ``Binarized
  neural networks: Training deep neural networks with weights and activations
  constrained to+ 1 or-1,'' \emph{arXiv preprint arXiv:1602.02830}, 2016.

\bibitem{rastegari2016xnor}
M.~Rastegari, V.~Ordonez, J.~Redmon, and A.~Farhadi, ``Xnor-net: Imagenet
  classification using binary convolutional neural networks,'' in \emph{Proc.
  ECCV}.\hskip 1em plus 0.5em minus 0.4em\relax Springer, 2016, pp. 525--542.

\bibitem{bocquet2018}
M.~Bocquet, T.~Hirztlin, J.-O. Klein, E.~Nowak, E.~Vianello, J.-M. Portal, and
  D.~Querlioz, ``In-memory and error-immune differential rram implementation of
  binarized deep neural networks,'' in \emph{IEDM Tech. Dig.}\hskip 1em plus
  0.5em minus 0.4em\relax IEEE, 2018, p. 20.6.1.

\bibitem{ly2018role}
D.~R.~B. Ly, A.~Grossi, C.~Fenouillet-Beranger, E.~Nowak, D.~Querlioz, and
  E.~Vianello, ``Role of synaptic variability in resistive memory-based spiking
  neural networks with unsupervised learning,'' \emph{J. Phys. D: Applied
  Physics}, 2018.

\bibitem{zhang2017130nm}
F.~Zhang, D.~Fan, Y.~Duan, J.~Li, C.~Fang, Y.~Li, X.~Han, L.~Dai, C.~Chen,
  J.~Bi \emph{et~al.}, ``A 130nm 1mb hfox embedded rram macro using
  self-adaptive peripheral circuit system techniques for 1.6 x work temperature
  range,'' in \emph{Proc. A-SSCC}.\hskip 1em plus 0.5em minus 0.4em\relax IEEE,
  2017, pp. 173--176.

\bibitem{ando2017brein}
K.~Ando, K.~Ueyoshi, K.~Orimo, H.~Yonekawa, S.~Sato, H.~Nakahara, M.~Ikebe,
  T.~Asai, S.~Takamaeda-Yamazaki, T.~Kuroda \emph{et~al.}, ``Brein memory: A
  13-layer 4.2 k neuron/0.8 m synapse binary/ternary reconfigurable in-memory
  deep neural network accelerator in 65 nm cmos,'' in \emph{Proc. VLSI Symp. on
  Circuits}.\hskip 1em plus 0.5em minus 0.4em\relax IEEE, 2017, pp. C24--C25.

\bibitem{sun2018fully}
X.~Sun, X.~Peng, P.-Y. Chen, R.~Liu, J.-s. Seo, and S.~Yu, ``Fully parallel
  rram synaptic array for implementing binary neural network with (+ 1,- 1)
  weights and (+ 1, 0) neurons,'' in \emph{Proc. ASP-DAC}.\hskip 1em plus 0.5em
  minus 0.4em\relax IEEE Press, 2018, pp. 574--579.

\bibitem{sun2018xnor}
X.~Sun, S.~Yin, X.~Peng, R.~Liu, J.-s. Seo, and S.~Yu, ``Xnor-rram: A scalable
  and parallel resistive synaptic architecture for binary neural networks,''
  \emph{algorithms}, vol.~2, p.~3, 2018.

\bibitem{tang2017binary}
T.~Tang, L.~Xia, B.~Li, Y.~Wang, and H.~Yang, ``Binary convolutional neural
  network on rram,'' in \emph{Proc. ASP-DAC}.\hskip 1em plus 0.5em minus
  0.4em\relax IEEE, 2017, pp. 782--787.

\bibitem{grossi2016fundamental}
A.~Grossi, E.~Nowak, C.~Zambelli, C.~Pellissier, S.~Bernasconi, G.~Cibrario,
  K.~El~Hajjam, R.~Crochemore, J.~Nodin, P.~Olivo \emph{et~al.}, ``Fundamental
  variability limits of filament-based rram,'' in \emph{IEDM Tech. Dig.}\hskip
  1em plus 0.5em minus 0.4em\relax IEEE, 2016, pp. 4--7.

\bibitem{gregori2003chip}
S.~Gregori, A.~Cabrini, O.~Khouri, and G.~Torelli, ``On-chip error correcting
  techniques for new-generation flash memories,'' \emph{Proc. IEEE}, vol.~91,
  no.~4, pp. 602--616, 2003.

\bibitem{chang201419}
M.-F. Chang, J.-J. Wu, T.-F. Chien, Y.-C. Liu, T.-C. Yang, W.-C. Shen, Y.-C.
  King, C.-J. Lin, K.-F. Lin, Y.-D. Chih \emph{et~al.}, ``19.4 embedded 1mb
  reram in 28nm cmos with 0.27-to-1v read using swing-sample-and-couple sense
  amplifier and self-boost-write-termination scheme,'' in \emph{Proc.
  ISSCC}.\hskip 1em plus 0.5em minus 0.4em\relax IEEE, 2014, pp. 332--333.

\bibitem{su2017462gops}
F.~Su, W.-H. Chen, L.~Xia, C.-P. Lo, T.~Tang, Z.~Wang, K.-H. Hsu, M.~Cheng,
  J.-Y. Li, Y.~Xie \emph{et~al.}, ``A 462gops/j rram-based nonvolatile
  intelligent processor for energy harvesting ioe system featuring nonvolatile
  logics and processing-in-memory,'' in \emph{Proc. Symp. VLSI Technol.}\hskip
  1em plus 0.5em minus 0.4em\relax IEEE, 2017, pp. T260--T261.

\bibitem{sassine2018sub}
G.~Sassine, C.~Nail, L.~Tillie, D.~A. Robayo, A.~Levisse, C.~Cagli, K.~E.
  Hajjam, J.-F. Nodin, E.~Vianello, M.~Bernard \emph{et~al.}, ``Sub-pj
  consumption and short latency time in rram arrays for high endurance
  applications,'' in \emph{Proc. IRPS}.\hskip 1em plus 0.5em minus 0.4em\relax
  IEEE, 2018, pp. P--MY.

\bibitem{whatmough2018dnn}
P.~N. Whatmough, S.~K. Lee, D.~Brooks, and G.-Y. Wei, ``Dnn engine: A 28-nm
  timing-error tolerant sparse deep neural network processor for iot
  applications,'' \emph{IEEE Journal of Solid-State Circuits}, vol.~53, no.~9,
  pp. 2722--2731, 2018.

\bibitem{lecun1998gradient}
Y.~LeCun, L.~Bottou, Y.~Bengio, and P.~Haffner, ``Gradient-based learning
  applied to document recognition,'' \emph{Proc. IEEE}, vol.~86, no.~11, pp.
  2278--2324, 1998.

\bibitem{krizhevsky2009learning}
A.~Krizhevsky and G.~Hinton, ``Learning multiple layers of features from tiny
  images,'' Citeseer, Tech. Rep., 2009.

\bibitem{abadi2016tensorflow}
M.~Abadi, P.~Barham, J.~Chen, Z.~Chen, A.~Davis, J.~Dean, M.~Devin,
  S.~Ghemawat, G.~Irving, M.~Isard \emph{et~al.}, ``Tensorflow: a system for
  large-scale machine learning.'' in \emph{OSDI}, vol.~16, 2016, pp. 265--283.

\end{thebibliography}
\bibliographystyle{IEEEtran}

\end{document}